# Polarization studies with NuSTAR

Simone Lotti[*a], Lorenzo Natalucci[a], Paolo Giommi[b], B. Grefenstette[c], Fiona A. Harrison[c], Kristin K. Madsen[c], Matteo Perri[b], Simonetta Puccetti[b] A. Zoglauer[d]

[a]INAF-Istituto di Astrofisica e Planetologia Spaziali, Roma, Italy; [b]ASI Science Data Center, Frascati, Italy; [c]Caltech Division of Physics, Mathematics and Astronomy, Pasadena, USA; [d]Space Sciences Laboratory, University of California at Berkeley, USA

**ABSTRACT**

The capability of NuSTAR to detect polarization in the Compton scattering regime (>50 keV) has been investigated. The NuSTAR mission, flown on June 2012 a Low Earth Orbit (LEO), provides a unique possibility to confirm the findings of INTEGRAL on the polarization of cosmic sources in the hard X-rays. Each of the two focal plane detectors are high resolution pixellated CZT arrays, sensitive in the energy range ~ 3 - 80 keV. These units have intrinsic polarization capabilities when the proper information on the double events is transmitted on ground. In this case it will be possible to detect polarization from bright sources on timescales of the order of $10^5$ s

## 1. INTRODUCTION

Several astrophysical processes are able to produce polarized photons: cyclotron emission (below 100 keV), synchrotron (with a polarization degree depending on the power-law index), curvature radiation, bremmstrahlung, and Compton effect. Measuring the polarization degree of a source, along with the spectral shape and timing information is a powerful tool to study the emission mechanisms responsible for the high energy emission. In particular the role of magnetic fields in a variety of objects, from accreting black holes in stellar mass systems and AGNs to isolated neutron stars, can be efficiently studied. Polarimetry can be also a probe of fundamental physics when observed from sources at cosmological distances[1].

While the detection of polarimetry below ~ 10 keV requires dedicated instrumentation, in the Compton scattering regime standard astronomical imaging detectors can provide measurements of polarization. This has been demonstrated by the recent measurement by INTEGRAL of polarized emission from the Crab[2] with a polarization degree of 46 ± 10%, and by the analysis of IBIS data from GRB041219A[3] and Cyg X-1[4]. These results, however, need further confirmation due to the large uncertainties and the high level of systematics affecting the instrument response to multiple events.

Within the next decade only two hard X-ray missions can obtain significant measurements of polarization from bright sources: NuSTAR[5] and Astro-H. On Astro-H, the SGD detector is sensitive to polarization in the range 50 – 200 keV with an effective area for polarized emission of ~ 30 cm$^2$ and a quality factor of about 80 %[6]. The NuSTAR detectors are sensitive in the energy range 3 - 80 keV and consist of two pixellated detector arrays in the focus of two multilayer optical mirrors[7]. Each array is built by four CZT crystals having a thickness of 2 mm, each equipped with a 32x32 ASIC readout. The electric design of the detectors result in a position resolution of 0.6 mm. Despite their low efficiency to Compton scattering, they benefit from a very low background. In the following we evaluate the sensitivity of NuSTAR in the detection of polarized radiation, using detailed simulations and estimates of the losses due to the positional detector response.

*Simone.lotti@iaps.inaf.it; phone 0039 06 4993 4690;

## 2.    MODELING OF POLARIZATION

In this study we make use of simulations that model the interaction of polarized and unpolarized photons in the CZT detector. The simulations are obtained with the 9.4 version of Geant4. The detector geometry is implemented as a 2 mm thick CZT block divided into 32 x 32 pixels (625 x 625 μm$^2$ each), for a total area of 4 cm$^2$. The energy threshold for a detection in each pixel is set at 3 keV. The method used is quite standard and is based on the analysis of multiple events in the NuSTAR detectors. The events useful for polarization are those produced by Compton scattering with escape of the secondary photon from the pixel volume of the primary interaction. One of the main limitations to the NuSTAR polarization sensitivity is the that the high throughput expected from double events in adjacent pixels is generated not only by absorption of the Compton scattered photons, but also by the charge sharing effect[8]. This effects the spatial distribution of the signal among the different pixels both for Compton scattering and photoelectric absorption. Therefore we have used a conservative approach: to consider only the double events with signals in non-adjacent pixels.

Since the mean free path of the secondary photon is very low in CZT in the operational range of NuSTAR, only a small fraction of events contribute to the polarization sensitivity. The azimuthal distribution of double events is a double peaked function that can be expressed as

$$\frac{dS}{d\phi} = \frac{S}{2\pi}\left[1 - QP\cos 2(\phi - \eta)\right] \quad (1)$$

where $P$ is the fraction of polarized photons, $\eta$ is the polarization angle (that corresponds to the minimum of the distribution) and $S$ is the source count rate. $Q$ is the modulation quality factor that represents the ratio between the amplitude of the modulation and the average level of the angle distribution. This depends on the azimuthal angle probability given by the Klein-Nishina cross-section but also on the actual geometry of the detector. The dependence of the $Q$ factor on the Compton scattering angle is dependent on photon energy (see Fig. 4.3 in Lei et al. 1997).

Near and below 100 keV the $Q$ factor peaks for scattering at right angles. This is optimum to the design of the NuSTAR CZT detector, for which the efficiency to detect scattered Compton photons is maximized at 90° (lower escape path of the secondary photon). On the other hand, the sensitivity is limited by the low cross-section for Compton scattering at E<80 keV and the charge sharing effect, that makes the signal of a single interaction be spread on two or more nearby pixels.

We have performed a number of simulations with the Geant4 package using either polarized and non-polarized photon beams, in order to investigate the sensitivity of the NuSTAR detector to detect polarization signals. We attempt to estimate the Minimum Detectable Polarization (MDP) as[9]

$$MDP = \frac{n_\sigma}{QS}\sqrt{\frac{S+B}{T}} \quad (2)$$

where $B$ is the background count rate (cts/s), $n_\sigma$ the desired S/N ratio and $T$ the exposure time. Note that $B$ is not the total background rate but the rate after the event selection. The aim of the simulations is to determine $Q$ and provide an estimate of the MDP via the above equation as a function of time and source intensity.

## 3.    DETECTOR RESPONSE TO POLARIZATION

Our first goal is to determine the energy range useful for polarization detection. In a first set of simulations we generated monochromatic beams with 10$^6$ photons perpendicular to the surface of the detector, uniformly distributed over the area of the central pixel. Then we select the double events for which the output signal exceeds the low energy threshold. Due to the CZT charge sharing effect we exclude the events with detections from the 8 pixels surrounding the central one. We performed several runs with monochromatic beams in the 48 - 78 keV energy range, for both unpolarized and polarized radiation. The direction of the polarization vector is chosen along the x-axis of detector. We then generated a detector map from each simulation of a polarized incident beam. On each map we performed a chi-square test to evaluate compatibility with a model map built by the corresponding simulations with unpolarized radiation. This allowed us to determine that the detector starts to be sensitive to polarization at ~50 keV: the compatibility is 57% at 48 keV, 1% at 53 keV. The simulations also show that in the energy band of interest for the polarization detection, it is easy to distinguish between the signal caused by the incident photon, and the one caused by the scattered one.

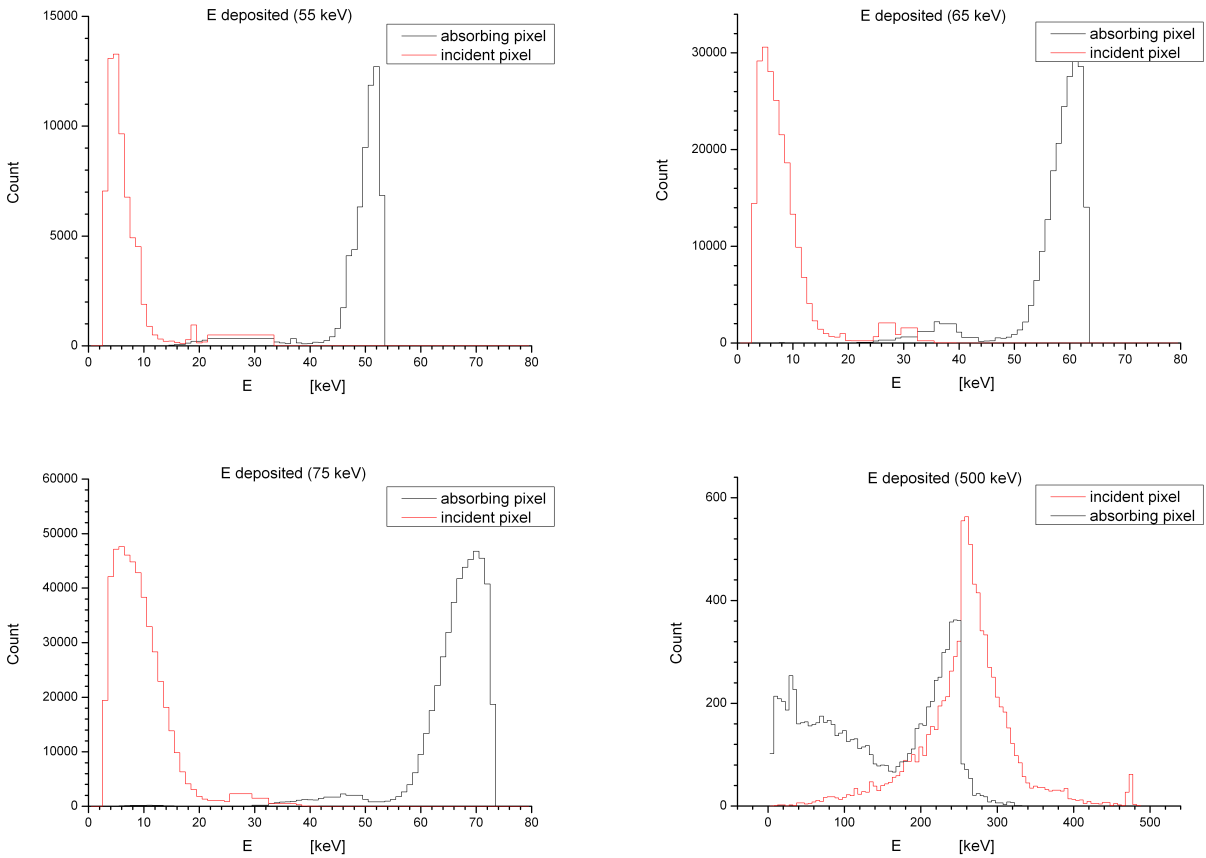

Fig. 1. Deposited Energy distribution for several incident beam energies. The red line represent the energy deposited in the Compton scattering event, while the black one is the energy deposited by the absorption of the scattered photon. In the energy band the signal in the absorber pixel is significantly higher than the one recorded in the absorber one.

In Fig. 1 are shown some distributions of the energy deposited in the central pixel (where the primary interaction occurs) and in the outer pixel (where the scattered photon is absorbed). The two distributions are well separated in the energy range 48-80 keV. This would not be the case at higher energies, as is shown for 500 keV. This means that we can recognize quite safely the incident photon pixel as the one which has the lowest energy deposit. Finally, we built for each simulation an individual azimuthal profile and fitted it by using expression (2). In this way we could estimate the predicted modulation factor as a function of energy (see Fig. 2).

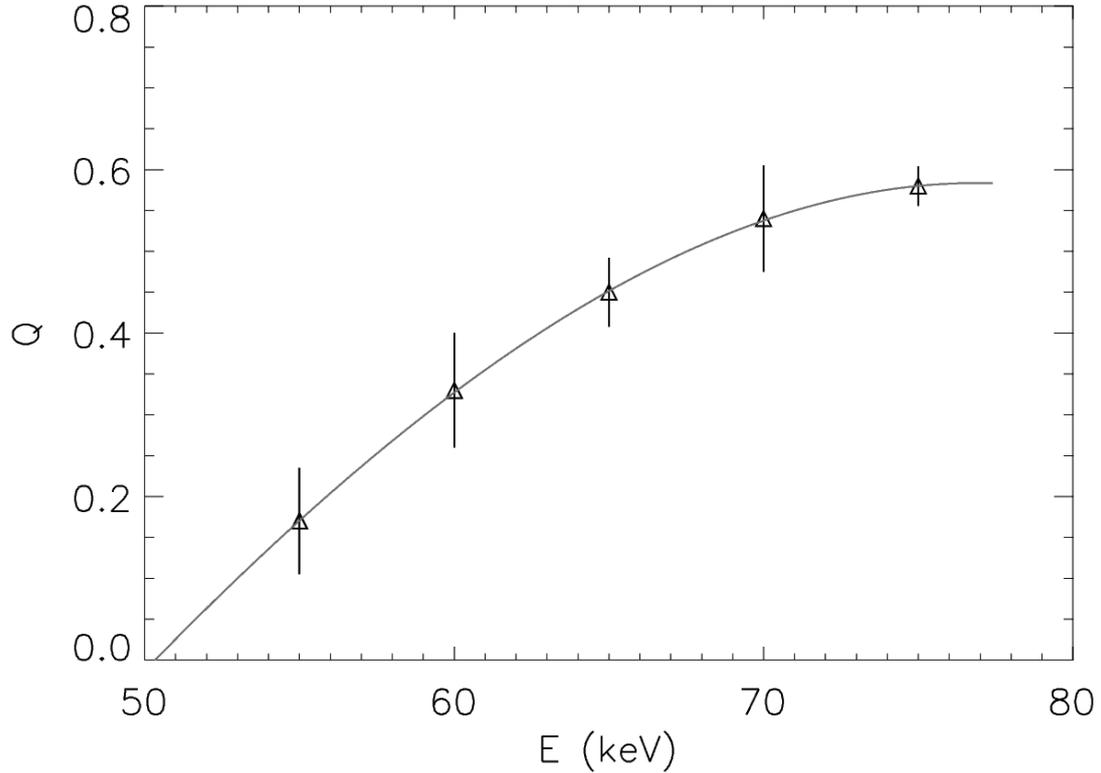

Fig. 2. The dependence of the quality factor Q on energy for $10^6$ polarized incident photons in polarization mode. The incident radiation is polarized along the grid lines of the detector.

## 4. BACKGROUND

To evaluate the background level for polarization mode we have used a full model of the NuSTAR satellite that has been readapted for GEANT4 simulations. We assumed a typical LEO environment impacting on the geometrical model of the satellite and processed the output of the detectors. The background of NuSTAR has been already described for the imaging mode[10], and we have used that reference spectrum to validate our own background simulations.

We then used our simulation to retrieve the rate of double events generated by the background particles in LEO. The ratio of double/single events background rate obtained with the simulations with no corrections is ~ 0.06. This corresponds to a rate of $2.5 \times 10^{-4}$ cts cm$^{-2}$ s$^{-1}$ in the 55 - 80 keV band. We have verified that the impact of this background level on the MDP is found to be less than 1% for sources brighter than 100 mCrab.

However the unrejected background of double events strongly depends on the conditions used to discriminate the events. There are several corrections that can be applied to reduce this level: first of all from energetic considerations we expect the energy deposited in the incident pixel not to exceed 20 keV, and the energy in the second pixel not being below 50 keV (see Fig. 1); we can also reject double events in which the first interaction occurs outside the PSF. Applying this conditions we have no events detected in several tens of minutes of simulated time. The evaluation of the true polarization background requires huge computational time. We therefore use the above conservative rate obtained for the total double events with no corrections.

## 5. SIMULATIONS OF CRAB NEBULA OBSERVATIONS

In order to investigate the sensitivity of the CZT detector to polarization, we decided to simulate two observations of the Crab Nebula corresponding to an exposure time of $10^5$ and $10^6$ seconds, with an input spectrum shape given by a power-law and photon energies ranging in the 55 - 80 keV. The spectral parameters assumed are a spectral index $\Gamma=2.136$ and a normalization N (1 keV) = 9.8 [11]. Firstly, we ran a 100% polarized beam to determine the modulation factor $Q$ associated with the Crab spectrum. The $Q$ value, obtained by fitting expression (1) to the azimuthal profile, is found to be 0.51±0.022, a value compatible with the values reported in Fig. 2. The double events distribution and the fit result is shown in figure 3.

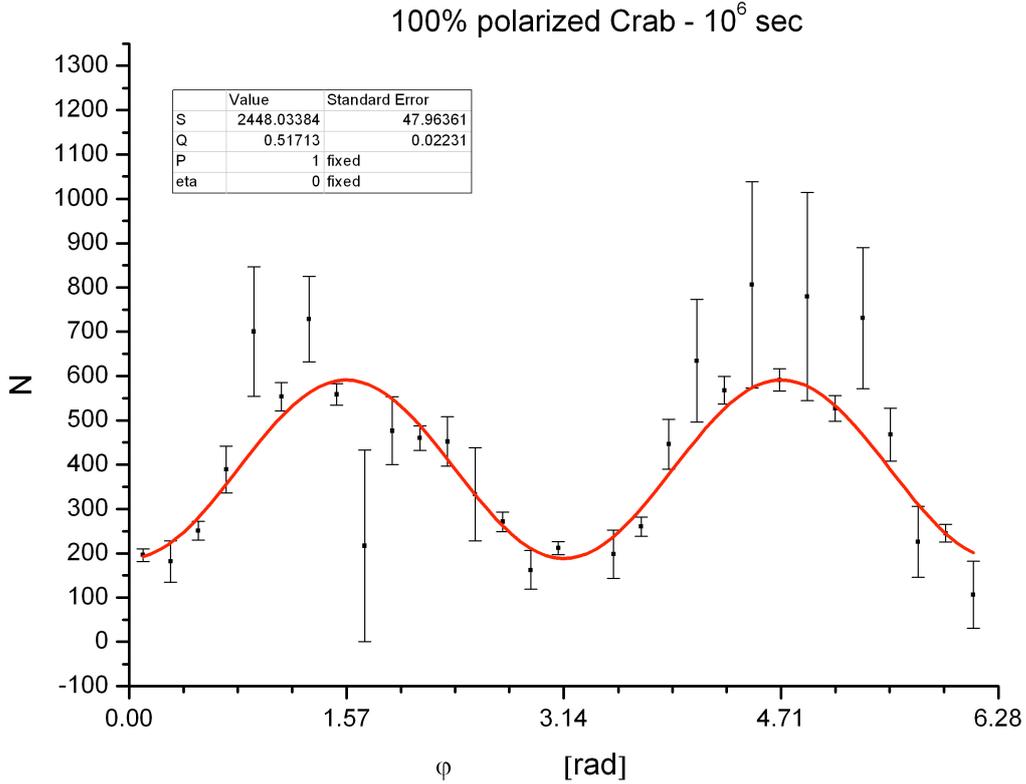

Fig. 3. Azimuthal distribution of double events resulting from a simulation of $10^6$ seconds fully polarized Crab nebula and the fit obtained with function (1)

Secondly, we simulated 4 observations of the Crab in the 55-78 keV corresponding to an observation time of ~ 1 Ms (~ 3 x $10^6$ photons, 46% linear polarization[12]). For each simulation we used a different value of the angle $\varphi$ between the source polarization vector and the axis orientation. In the following we refer to this angle as polarization angle. The angles of polarization were set to 0°, 15°, 30° and 45° respect to the X axis of the instrument. In table 1 are shown the values of polarization fraction and angle recovered fitting the data with (1) for the four cases analyzed using the $Q$ = 0.51 value obtained before. For each event, the incidence pixel is determined as the one with the lower energy deposit.

Table 1. Values obtained fitting simulated data with $Q = 0.51$

| Expected φ | φ | P |
|---|---|---|
| 0° | 0 ± 3.8° | 0.46 ±0.05 |
| 15° | 11.1° ±4° | 0.43 ±0.05 |
| 30° | 25.8° ± 3.8° | 0.45 ±0.06 |
| 45° | 44.7° ± 4.2° | 0.40 ±0.07 |

A simulation with unpolarized photons was also analyzed to confirm the correct behavior of the program and of the analysis method and reported a result consistent with an unpolarized beam with a reduced $\chi^2 = 0.9$.

## 6. RESULTS

Using our evaluation of the quality factor $Q$ and the knowledge of the detector background, one should be able to assess the MDP as a function of the source intensity. Results on the NuSTAR background evaluation have been reported in Harrison et al.[5]. However, the background rates and spectra of use in the polarization analysis are very different from the total detected background, as we select only double events and apply selection rules pertaining to the probability of the primary/secondary events locations and energies. We used the data reported in the literature to validate our background simulations for imaging mode, and then calculated the expected background for polarization measurements. In any case the results are consistent with a negligible contribute of the background to the polarization sensitivity. In figure 4 is shown the MDP curve as a function of source intensity, for observation times of 1 Ms and 100 ks.

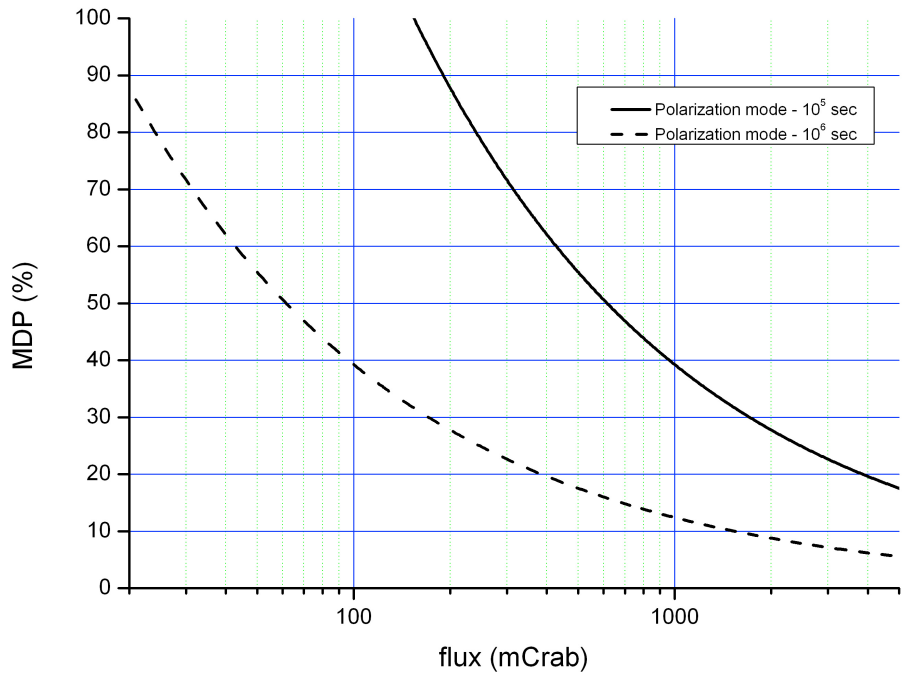

Fig. 4. Minimum Detectable Polarization as function of source flux for an exposure time of $10^5$ s and $10^6$ s, 5σ detection

From these preliminary results, it comes out that NuSTAR should be able to improve the current observation of the Crab polarization since the first in-flight calibration campaign.

## 7. CONCLUSIONS

We have performed simulations to estimate the capability of NuSTAR to detect polarization of the cosmic hard X-ray sources. The results obtained show that the NuSTAR CZT detectors have a sensitivity for polarization measurements better than the currently flying hard X-ray satellites, like INTEGRAL.

The results shown in Table 1, for measurements of the polarization fraction are quite compatible with the MDP curve calculated theoretically on the basis of the expected background and effective area. In particular, we have shown that the polarization in the Crab Nebula should be detected in observation times of the order of 100 ks. Moreover, due to the very low background the polarization signal can be detected for observation times of the order of $10^6$ s. The final sensitivity actually depends on the source spectrum and will be better for sources harder than the Crab, for example, bright accreting black holes like Cyg X-1.

Unfortunately, the NuSTAR science telemetry does not currently transmit the full information of the multiple events (position and energy). Hence a polarization analysis based on the described approach cannot be applied for the present flight configuration. The opportunity to upgrade the instrument's software to provide the full information on these events is currently being evaluated by the Team.

**References**

[1] Y.Z. Fan, D. Wei & D. Xu 2007, γ-ray burst ultraviolet/optical afterglow polarimetry as a probe of

quantum gravity , MNRAS, 376, 1857.

[2] A.J. Dean et al. 2008, Polarized Gamma-Ray Emission from the Crab , Science, 321, 1883

[3] D.Gotz et al. 2009, Variable polarization measured in the prompt emission of GRB 041219A using

IBIS on board INTEGRAL , Ap.J Lett., 695, L208

[4] P. Laurent et al. 2011, Polarized Gamma-Ray Emission from the Galactic Black Hole Cygnus X-1 ,

Science, 332, 438

[5] F.A. Harrison et al. 2010, The Nuclear Spectroscopic Telescope Array (NuSTAR) , Proc. SPIE Vol.

7732, 77320S

[6] H. Tajima et al. 2010, Polarimetry with ASTRO-H soft gamma-ray detector , "X-ray Polarimetry: A

New Window in Astrophysics" by Ronaldo Bellazzini, Enrico Costa, Giorgio Matt and Gianpiero

Tagliaferri. Cambridge University Press, p.275

[7] V. R. Rana et al. 2009, "Development of Focal Plane Detectors for the Nuclear Spectroscopic Telescope Array (NuSTAR) Mission", SPIE, 7435, 2.

[8] Hills, G.L. 1997, Space Sci. Rev. 82, 309 Rana, V.R. et al. 2009, Proc. SPIE Vol. 7435, 743503-1

[9] Lei, F., Dean, A.J., Hills, G.L.: Space Sci. Rev. 82, 309–388, (1997)

[10] F.Harrison, K. Madsen - NuSTAR: instrument performance guide -

http://www.NuSTAR.caltech.edu/uploads/files/NuSTAR_performance_v1.pdf

[11] Kirsch et al, 2005 "Crab: the standard X-ray candle with all modern x-ray satellites" - Proceedings of the SPIE, Volume 5898, pp. 22-33 - DOI: 10.1117/12.616893

[12] A. J. Dean et al, Science 29 August 2008: Vol. 321 no. 5893 pp. 1183-1185 DOI: 10.1126/science.1149056